\documentclass[aps,pre,reprint,twocolumn,showpacs,preprintnumbers,groupedaddress,floatfix,amsmath,amssymb]{revtex4-1}

\usepackage{graphicx}
\usepackage{color}

\newcommand{\physrep}{Phys.~Rep.} 
\newcommand{\apjl}{Astrophys.~J.~Lett.} 
\newcommand{\aap}{Astron.~\&~Astrophys.} 
\newcommand{\araa}{Annu.~Rev.~Astron.~Astrophys.} 
\newcommand{\mnras}{Mon.~Not.~R.~Astron.~Soc.} 
\newcommand{\ssr}{Space Science Rev.} 
\newcommand{\pasp}{Publ.~Astron.~Soc.~Pac.} 

\newcommand{\boldnabla}{\mbox{\boldmath$\nabla$}}
\newcommand{\tabhe}{\parbox[0pt][2em][c]{0cm}{}}

\begin{document}


\title{Magnetic Field Amplification by Small-Scale Dynamo Action: \\ Dependence on Turbulence Models and Reynolds and Prandtl Numbers}

\author{Jennifer Schober}
 \email{schober@stud.uni-heidelberg.de}
 \affiliation{Institut f\"ur Theoretische Astrophysik, Zentrum f\"ur Astronomie der Universit\"at Heidelberg, Albert-\"Uberle-Strasse\ 2, D-69120 Heidelberg, Germany}
\author{Dominik Schleicher}
  \email{dschleic@astro.physik.uni-goettingen.de}
  \affiliation{Institut f\"ur Astrophysik, Georg-August-Universit\"at G\"ottingen, Institut f\"ur Astrophysik, Friedrich-Hund-Platz, D-37077 G\"ottingen, Germany}
\author{Christoph Federrath}
  \email{federrath@uni-heidelberg.de}
  \affiliation{Monash Centre for Astrophysics (MoCA), School of Mathematical Sciences, Monash University, Victoria 3800, Australia;\\
               CRAL, Ecole Normale Sup\'{e}rieure de Lyon, F-69364 Lyon, France; and\\ 
               Institut f\"ur Theoretische Astrophysik, Zentrum f\"ur Astronomie der Universit\"at Heidelberg, Albert-\"Uberle-Strasse\ 2, D-69120 Heidelberg, Germany}
\author{Ralf Klessen}
  \email{klessen@uni-heidelberg.de}
  \affiliation{Institut f\"ur Theoretische Astrophysik, Zentrum f\"ur Astronomie der Universit\"at Heidelberg, Albert-\"Uberle-Strasse\ 2, D-69120 Heidelberg, Germany }
\author{Robi Banerjee}
  \email{banerjee@hs.uni-hamburg.de}
  \affiliation{Hamburger Sternwarte, Gojenbergsweg 112, D-21029 Hamburg, Germany}

\date{\today}

\begin{abstract}
The small-scale dynamo is a process by which turbulent kinetic energy is converted into magnetic energy, and thus it is expected to depend crucially on the nature of the turbulence. In this paper, we present a model for the small-scale dynamo that takes into account the slope of the turbulent velocity spectrum $v(\ell) \propto \ell^{\vartheta}$, where $\ell$ and $v(\ell)$ are the size of a turbulent fluctuation and the typical velocity on that scale. The time evolution of the fluctuation component of the magnetic field, i.e., the small-scale field, is described by the Kazantsev equation. We solve this linear differential equation for its eigenvalues with the quantum-mechanical WKB-approximation. The validity of this method is estimated as a function of the magnetic Prandtl number Pm. We calculate the minimal magnetic Reynolds number for dynamo action, $\text{Rm}_{\text{crit}}$, using our model of the turbulent velocity correlation function. For Kolmogorov turbulence ($\vartheta=1/3$), we find that the critical magnetic Reynolds number is $\text{Rm}_{\text{crit}}^\text{K} \approx 110$ and for Burgers turbulence ($\vartheta=1/2$) $\text{Rm}_{\text{crit}}^\text{B} \approx 2700$. Furthermore, we derive that the growth rate of the small-scale magnetic field for a general type of turbulence is $\Gamma \propto \text{Re}^{(1-\vartheta)/(1+\vartheta)}$ in the limit of infinite magnetic Prandtl numbers. For decreasing magnetic Prandtl number (down to $\text{Pm}\gtrsim10$), the growth rate of the small-scale dynamo decreases. The details of this drop depend on the WKB-approximation, which becomes invalid for a magnetic Prandtl number of about unity.
\end{abstract}

\maketitle

\section{\label{sec:Introduction} Introduction}

Magnetic fields are observed in the whole Universe on different length scales and they play an important role for virtually all astrophysical objects. To probe the origin of these fields, it is crucial to gain better theoretical insight into their dynamical properties. \\ 
Observations suggest that magnetic fields were already space filling in the early Universe. Hints toward high-redshift magnetic fields come from the study of Bernet\textit{ et al.\ }\cite{Bernet2008}. Recent studies using $\gamma$-ray observations of the intergalactic medium find a lower limit of $B_0\approx10^{-8}$--$10^{-6}~\text{nG}$ \cite{Tavecchio2010,NeronovVovk2010,Taylor2011}, indicative of an early generation scenario. For the very early Universe, we have only upper limits, but no direct observations showing the presence of magnetic fields. Yamazaki\textit{ et al.\ }\cite{Yamazaki2006} derive an upper limit of the magnetic field strength from the cosmic microwave background temperature anisotropy. They predict that the magnetic field strength at the present scale of 1 Mpc \footnote{$1~\text{Mpc} \approx 3.1\times10^{22}~\text{m}$} is $B_0 \lesssim 4.7~\text{nG}$. Moreover, Schleicher and Miniati \cite{SchleicherMiniati2011} present a method to derive the magnetic field strength in the reionization epoch (redshift $z>6$). They show that the upper halo mass is set by the magnetic pressure and conclude, by using data from the reionization epoch, that the co-moving magnetic field strength is $B_0 \lesssim 3~\text{nG}$ (see also \cite{Schleicher2009}). \\
Besides these observational constraints, there are different theories that describe the origin of these primordial magnetic fields. The first seed fields could already have been produced during inflation. Turner and Widrow \cite{TurnerWidrow1988} find that $B_0\approx10^{-25}$--$10^{-1}$ nG on a scale of $1$ Mpc can be produced when the conformal invariance is broken. Following Sigl\textit{ et al.\ }\cite{Sigl1997}, there is also the possibility of creating a magnetic field during the phase transitions in the very early Universe. They predict a field strength $B_0 \approx 10^{-20} ~\text{nG}$ from the electroweak phase transition and $B_0 \approx 10^{-11} ~\text{nG}$ from the quantum chromodynamics (QCD) phase transition. \\
The observed magnetic field strengths in high-redshift as well as in present-day galaxies are, however, typically orders of magnitude higher than predicted by the theories. For example the Milky Way has locally a magnetic field of $(6\pm2)\times10^3 ~\text{nG}$ \cite{Beck2001}. Hence there must be additional and strong amplification. The simplest way of amplifying a magnetic field by gravitational compression under the condition of flux freezing, when the magnetic field is perfectly coupled to the gas. With the conservation of magnetic flux, one can directly show that $B\propto \rho^{2/3}$ for spherical collapse, where $\rho$ is the mass density of the gas. However, magnetohydrodynamical dynamos can amplify small initial magnetic fields exponentially in time, leading to a growth of $B$ much faster than $\rho^{2/3}$ \cite{Sur2010,Federrath2011.1, Schleicher2010}. A dynamo, in general, is a process by which kinetic energy is converted into magnetic energy. One has to distinguish two different types of dynamo: the \textit{large-scale dynamo}, which is excited by large-scale motions, and the \textit{small-scale dynamo}, which is excited by turbulence on very small scales. Dynamos amplify the magnetic energy exponentially. The small-scale dynamo is important in particular during the formation of the first stars and galaxies, where turbulence is released from the gravitational potential well, providing the means to amplify magnetic fields \cite{Schleicher2010}. The same processes may also occur during the formation of large-scale structure \cite{Ryu2008}. In most astrophysical objects, the growth rate of the small-scale dynamo is very high and that can explain the high magnetic field strengths observed in the primordial as well as in the present-day Universe. \\
The small-scale dynamo was first suggested by Batchelor in 1950 \cite{Batchelor1950}. In 1967, Kazantsev developed a spectral theory of this process \cite{Kazantsev1967,Kazantsev1968}. He derived a differential equation, the \textit{Kazantsev equation}, which describes the evolution of the magnetic energy of the small-scale field. This equation has been solved for different special cases. For example, Subramanian \cite{Subramanian1997} used this theory to describe the small-scale dynamo in the case of incompressible turbulence, i.e., Kolmogorov turbulence. In the limit of infinite magnetic Prandtl number Pm, which is the ratio of the magnetic to the hydrodynamical Reynolds numbers, Rm and Re, he found a growth rate of the small-scale magnetic energy of $(15V)/(4L)\sqrt{\text{Re}}$ \cite{BrandenburgSubramanian2005}. Schekochihin\textit{ et al.\ }\cite{Schekochihin2002} analyzed the Kazantsev theory for an arbitrary compressibility (and spatial dimension), also in the limit of large Prandtl number. However, they did not give a model for the correlation function of the turbulent velocity field, but used a small-scale approximation of this tensor. There are also alternative approaches for the limit of small Prandtl number based on the separation of scales \cite{RogachevskiiKleerin1997, KleeorinRogachevskii2011,ZeldovichRuzmaikinSokoloff1990}, which find similar solutions of the Kazantsev equation. Further ways to solve this equation exist, but go beyond the scope of this work (see for example \cite{KleeorinRogachevskiiSokoloff2002,KleeorinRogachevskii2008}). \\
In this paper, we aim to explore dynamo amplification for turbulence models with $v(\ell) \propto \ell^{\vartheta}$ in the inertial range, i.e., the range of the kinetic energy cascade. Here, $\ell$ is the size of a turbulent fluctuation and $v(\ell)$ the typical velocity on this scale. The range of $\vartheta$ goes from $1/3$, which describes the incompressible Kolmogorov turbulence \cite{Kolmogorov1941}, to $1/2$ for Burgers turbulence \cite{Burgers1948}. In astrophysical objects we expect that the exponent $\vartheta$ lies between those extrema \cite{SheLeveque1994,Boldyrev2002,Larson1981,Federrath2010,OssenkopfMacLow2002}. Burgers turbulence is often referred to as highly compressible turbulence, obtained in the limit of very high Mach numbers. \\
Federrath\textit{ et al.\ }\cite{Federrath2011.2} explore the Mach number dependence of the small-scale dynamo and show, that the dynamo works for both, subsonic and highly supersonic turbulence. Furthermore, these numerical experiments indicate that the small-scale magnetic field saturates after a certain time. They find that the saturation levels and growth rates depend strongly on the nature of the turbulence. However, most simulations typically have a magnetic Prandtl number of $\text{Pm}\approx1$, requiring further efforts and understanding for $\text{Pm}\gg1$ and $\text{Pm}\ll1$.\\
Here, we present an improved analytical model for the small-scale dynamo that takes into account highly compressible turbulence also. In Sec. \ref{sec: AnalyticFramework}, we outline the main ideas and equations behind the Kazantsev theory and describe the small-scale dynamo mathematically. We use the WKB approximation as a solution of the Kazantsev equation, which is valid in the limit of large magnetic Prandtl number. However, the evolution equation of the magnetic energy turns out to depend on the correlation function of the turbulent velocity field. Therefore, we present a general model for this correlation function for turbulence of arbitrary type with $v(\ell) \propto \ell^{\vartheta}$ in the inertial range. In Sec. \ref{sec: Results}, we calculate the critical magnetic Reynolds number for the small-scale dynamo, $\text{Rm}_{\text{crit}}$. Furthermore, we determine the growth rate $\Gamma$ of the magnetic energy. The latter is calculated in the limit $\text{Pm}\rightarrow\infty$ and for finite magnetic Prandtl numbers ($\text{Pm}\gtrsim10$). In Sec. \ref{sec: Discussion}, we discuss our results and compare them to the literature. We close our work with conclusions in Sec. \ref{sec: Con&Out}.

\section{\label{sec: AnalyticFramework} Analytic Framework}

\subsection{Kazantsev theory}

In 1967 Kazantsev set up a theory for describing the time evolution of the magnetic energy that grows due to turbulent motions of a conducting fluid \cite{Kazantsev1968}. The mechanism of converting kinetic energy into magnetic energy in this way is known as the turbulent or small-scale dynamo. In this section we describe the Kazantsev theory following mainly the formalism proposed by Brandenburg and Subramanian \cite{BrandenburgSubramanian2005} and Subramanian \cite{Subramanian1997}.

\subsubsection{The turbulent velocity field}
The theoretical description of turbulence starts with decomposition of the velocity field $\textbf{v}$ into a mean field $\left\langle\textbf{v}\right\rangle$ and a turbulent component $\delta\textbf{v}$:
\begin{equation}
  \textbf{v} = \left\langle\textbf{v}\right\rangle + \delta\textbf{v}.
\end{equation}
Following the work of Taylor \cite{Taylor1935} we model the spatial appearance of turbulence via the two-point correlation function. The correlation of two turbulent velocity components at the positions $\textbf{r}_1$ and $\textbf{r}_2$ at the times $t$ and $s$ for a Gaussian random velocity field with zero mean, which is isotropic, homogeneous and $\delta$-correlated in time, is
\begin{equation}
  \left\langle \delta v_i(\textbf{r}_1,t)\delta v_j(\textbf{r}_2,s)\right\rangle = T_{ij}(r)\delta(t-s)
\end{equation}
with the two-point correlation function $T_{ij}(r)$, where $r\equiv|\textbf{r}_1-\textbf{r}_2|$. It was shown by Batchelor \cite{Bachelor1953} that the correlation function can be divided into a transverse part $T_{\text{N}}$ and a longitudinal part $T_{\text{L}}$ in the following way:
\begin{equation}
  T_{ij}(r) = \left(\delta_{ij}-\frac{r_i r_j}{r^2}\right)T_{\text{N}}(r) + \frac{r_ir_j}{r^2}T_{\text{L}}(r).
\end{equation}
We neglect here the effect of helicity, which would appear as an additional term in $T_{ij}$.\\
In the special case of a divergence-free turbulent velocity field ($\text{div}~\delta \textbf{v} = 0$), characteristic of incompressible fluids, we find that the transverse correlation function is connected to the longitudinal one by
\begin{equation}
  T_{\text{N}}(r) = \frac{1}{2r} \frac{\text{d}}{\text{d} r}\left(r^2T_\text{L}(r)\right).
\label{TNdivfree}
\end{equation}
For the other extreme case, an irrotational turbulent velocity field ($\text{rot}~\delta \textbf{v} = 0$), as expected for purely shock-dominated flows, we find the relation
\begin{equation}
  T_{\text{L}}(r) = r\frac{\text{d}T_{\text{N}}(r)}{\text{d} r} + T_{\text{N}}(r).
\label{TNrotfree}
\end{equation}

\subsubsection{Kazantsev equation}
Like the velocity field, the magnetic field can be separated into a mean field $\left\langle\textbf{B}\right\rangle$ and a fluctuation part $\delta \textbf{B}$:
\begin{equation}
  \textbf{B} = \left\langle \textbf{B}\right\rangle + \delta \textbf{B}.
\end{equation}
Now let us assume that the fluctuating component $\delta \textbf{B}$, like the velocity field, is a homogeneous, isotropic Gaussian random field with zero mean. Then we can write the correlation function as
\begin{equation}
  \left\langle \delta B_i(\textbf{r}_1,t) \delta B_j(\textbf{r}_2,t)\right\rangle = M_{ij}(r,t)
\end{equation}
with the two-point correlation function
\begin{equation}
  M_{ij}(r,t) = \left(\delta_{ij}-\frac{r_ir_j}{r^2}\right)M_{\text{N}}(r,t) + \frac{r_ir_j}{r^2}M_{\text{L}}(r,t).
\label{correlationB}
\end{equation}
We will omit the dependencies for a better overview in most of the following equations. \\
As the magnetic field is always divergence-free, i.e., $\partial/\partial r_{1i}~M_{ij}(r,t) = \partial/\partial r_{1j}~M_{ij}(r,t) = 0$, we can derive a relation between the transverse and the longitudinal correlation function similar to (\ref{TNdivfree}):
\begin{equation}
  M_{\text{N}} = \frac{1}{2r} \frac{\text{d}}{\text{d} r}\left(r^2M_\text{L}\right),
\label{MN}
\end{equation}
where we have used that $(r_ir_j/r^2)M_{ij}=M_\text{L}$ and $(r_i/r_j)M_{ij}=M_\text{N}$. The time derivative of $\left\langle \delta B_i\delta B_j\right\rangle$ is
\begin{eqnarray}
  \frac{\partial M_{ij}}{\partial t} & = & \frac{\partial}{\partial t}\left(\left\langle \delta B_i\delta 																														               B_j\right\rangle \right) \nonumber \\
                                     & = & \frac{\partial}{\partial t}\left(\left\langle B_iB_j\right\rangle -  \left\langle                                                          B_i\right\rangle\left\langle B_j \right\rangle\right)  \nonumber \\
                                     & = & \left\langle\frac{\partial B_i}{\partial t}B_j\right\rangle + \left\langle B_i                                                             \frac{\partial B_j}{\partial t}\right\rangle \nonumber \\
                                     &   & -\frac{\partial}{\partial t}\left(\left\langle B_i\right\rangle\left\langle                                                                B_j\right\rangle\right).
  \label{DeriMij}
\end{eqnarray}
In the upper equation we can substitute the induction equation
\begin{equation}
  \frac{\partial \textbf{B}}{\partial t} = \boldnabla\times\textbf{v}\times\textbf{B} - \eta\boldnabla\times\boldnabla\times\textbf{B},
\label{induction}
\end{equation}
where $\eta \equiv c^2/(4\pi\sigma)$ is the magnetic diffusivity with the speed of light $c$ and the electrical conductivity $\sigma$, and the evolution equation of the magnetic mean field
\begin{equation}
  \frac{\partial\left\langle \textbf{B}\right\rangle}{\partial t} = \boldnabla \times\left[\left\langle \textbf{v}\right\rangle \times                                                                             \left\langle \textbf{B}\right\rangle -                                                                                                     \eta_\text{eff}\boldnabla\times\left\langle \textbf{B}\right\rangle\right]
\label{meanB}
\end{equation}
with the effective parameter $\eta_\text{eff}=\eta+T_\text{L}(0)$. After a lengthy derivation \cite{BrandenburgSubramanian2005} this leads to
\begin{eqnarray}
  \frac{\partial M_\text{L}}{\partial t} & = & 2\kappa_\text{diff} M_\text{L}'' 
                                               + 2\left(\frac{4\kappa_\text{diff}}{r}+ \kappa_\text{diff}'\right) M_\text{L}'                                                             \nonumber \\
                                         &   & + \frac{4}{r}\left(\frac{T_\text{N}}{r} - \frac{T_\text{L}}{r} - T_\text{N}'                                                                - T_\text{L}'\right) M_\text{L} 
\label{dMLdt}
\end{eqnarray}
with
\begin{equation}
  \kappa_\text{diff}(r) = \eta + T_\text{L}(0) - T_\text{L}(r).
\label{kappaN}
\end{equation}
The prime denotes differentiation with respect to $r$. The diffusion of the magnetic correlations, $\kappa_\text{diff}$, contains in addition to the magnetic diffusivity $\eta$ the scale-dependent turbulent diffusion $T_\text{L}(0) - T_\text{L}(r)$.\\
With the solution of Eq. (\ref{dMLdt}) we can calculate $M_\text{N}$ also by using the relation (\ref{MN}) and so find the total correlation function of the magnetic field fluctuations $M_{ij}$. We note that this quantity is proportional to the energy density of the magnetic field, $B^2/(8\pi)$.\\
In order to separate the time from the spatial coordinates we use the ansatz 
\begin{equation}
  M_\text{L}(r,t) \equiv \frac{1}{r^2\sqrt{\kappa_\text{diff}}}\psi(r)\text{e}^{2\Gamma t}.
\end{equation}
Substitution of this ansatz in Eq. (\ref{dMLdt}) gives us
\begin{equation}
  -\kappa_\text{diff}(r)\frac{\text{d}^2\psi(r)}{\text{d}^2r} + U(r)\psi(r) = -\Gamma \psi(r).
\label{Kazantsev}
\end{equation}
This is the \textit{Kazantsev equation}. It formally looks like the quantum-mechanical Schr\"odinger equation with a ``mass" $\hbar^2/(2\kappa_\text{diff})$ and the ``potential"
\begin{equation}
  U(r) \equiv \frac{\kappa_\text{diff}''}{2} - \frac{(\kappa_\text{diff}')^2}{4\kappa_\text{diff}} +                                                     \frac{2\kappa_\text{diff}}{r^2} + \frac{2T_\text{N}'}{r} + \frac{2(T_\text{L}-T_\text{N}+\kappa_\text{diff})}{r^2}.
\label{GeneralPotential}
\end{equation}
It describes the \textit{kinematic limit}, because $U$ is independent of the time.\\

\subsection{WKB approximation}
We can use common methods from quantum mechanics, like the WKB approximation, to solve the Kazantsev equation (\ref{Kazantsev}). WKB stands for Wentzel, Kramers, and Brillouin, who developed this method in 1926 \cite{Kramers1926}.

\subsubsection{Solution of the Kazantsev equation in the WKB approximation}
In order to use the standard WKB method, we have to make some substitutions. Let us first introduce a new radial coordinate $x$ by defining $r \equiv \text{e}^x$.
This leads to
\begin{eqnarray}
  \frac{\kappa_\text{diff}(x)}{\text{e}^x}\frac{\text{d}}{\text{d}x}\left(\frac{1}{\text{e}^x}\frac{\text{d}\psi(x)}{\text{d}x}\right)      - \left(\Gamma + U(x)\right)\psi(x)  =  0. 
\end{eqnarray}
Next we eliminate the first-derivative terms through the substitution
\begin{equation}
  \psi(x) \equiv \text{e}^{x/2}\theta(x),
\end{equation}
to obtain
\begin{equation}
  \frac{\text{d}^2\theta(x)}{\text{d}x^2} + p(x)\theta(x) = 0
\label{Kazantsev2}
\end{equation}
with the definition 
\begin{equation}
  p(x) \equiv  - \frac{[\Gamma + U(x)]\text{e}^{2x}}{\kappa_\text{diff}(x)} - \frac{1}{4}.
\label{p}
\end{equation}
The WKB solutions of Eq. (\ref{Kazantsev2}) are linear combinations of
\begin{equation}
  \theta(x) = \frac{1}{p^{1/4}}\text{exp}\left(\pm i \int\sqrt{p(x')}\text{d}x'\right).
\end{equation}
The boundary conditions for $\psi(r)$ and $\theta(x)$ are
\begin{eqnarray}
  \psi(r) & \xrightarrow{r \rightarrow 0,\infty} & 0       \nonumber  \\
  \Rightarrow \theta(x) & \xrightarrow{r \rightarrow \pm \infty} & 0.
\end{eqnarray}
We can make some predictions about the shape of the function $\theta(x)$. For very small $x$ ($x\rightarrow-\infty$), $p(x)$ goes to $-1/4 < 0$, which leads to exponentially growing and decaying solutions of $\theta$. In the other limit ($x\rightarrow\infty$), $p(x)\rightarrow-\Gamma/\eta ~ \text{e}^{2x}$, we have growing mode solutions only for positive $\Gamma$. The boundary conditions require that $\theta$ needs to grow exponentially for $x\rightarrow-\infty$ and decay exponentially at $x\rightarrow\infty$. In order to arrange this, $p(x)$ must go through zero, so $U(x)$ needs to become negative for some $r$. From now on we label the roots of $U(x)$ as $x_1$ and $x_2>x_1$. As $U(r)$ becomes negative for some $r$, $p(r)$ becomes positive for certain values of $r$. This means that we have oscillatory solutions for $x_1<x<x_2$. The condition for the eigenvalues $\Gamma$ in this case is \cite{MestelSubramanian1991}
\begin{equation}
  \int_{x_1}^{x_2}\sqrt{p(x')}\text{d}x' = \frac{2n+1}{2}\pi
\label{eigenvalue}
\end{equation}
for different excitation levels $n\in\mathbb{N}$. In this work we concentrate on the lowest mode $n=0$.

\subsubsection{Validity of the WKB approximation in general}
In order to find the limits in which the WKB method leads to valid solutions of the Kazantsev equation, we derive the differential equation that is solved exactly by 
\begin{equation}
  \theta(x) = \frac{1}{p^{1/4}}\text{exp}\left(\pm i \int_{x_1}^{x_2}\sqrt{p(x')}\text{d}x'\right).                      
\end{equation}
The second derivative of $\theta(x)$ with respect to $x$ can be written as
\begin{equation}
  \theta''(x) + \left(1 + \frac{p''}{4p^2} - \frac{3}{16}\frac{(p')^2}{p^3}\right)p\theta(x) = 0,                       
\end{equation}
where now the prime denotes $\text{d}/\text{d}x$. This equation results in the Kazantsev equation (\ref{Kazantsev2}) if
\begin{equation}
  \left| f(x) \right| \ll 1,       
\label{CheckApprox}               
\end{equation}
with
\begin{equation}
  f(x) \equiv \frac{p''}{4p^2} - \frac{3}{16}\frac{(p')^2}{p^3}.       
\end{equation}
We use this result in the Appendix to check the range of parameters in which the WKB method produces accurate solutions of the Kazantsev equation.

\subsection{\label{sec: Spectrum} Modeling the turbulent correlation function}

\begin{figure}
  \includegraphics[width=0.48\textwidth]{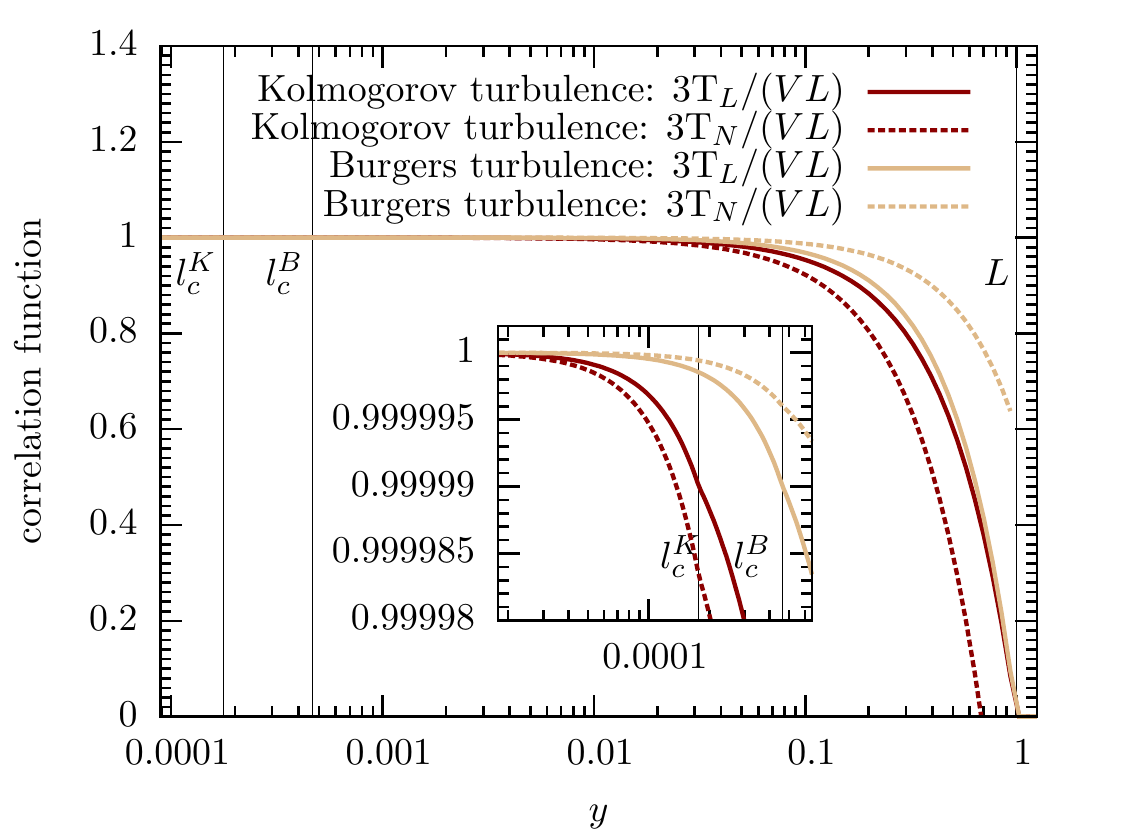}
  \caption{(Color online) Dependence of the longitudinal and transverse correlation functions $T_\text{L}$ and $T_\text{N}$ on the dimensionless parameter $y\equiv r/L$ for Kolmogorov ($\vartheta=1/3$) and Burgers ($\vartheta=1/2$) turbulence. We choose a fixed Reynolds number of $10^5$. The vertical lines indicate the cutoff scale of the turbulence $\ell_\text{c}$ and the largest scale of the eddies $L$. Notice that the cutoff scale for Kolmogorov turbulence ($\ell_\text{c}^\text{K}=\text{Re}^{-3/4}L$) is different from the one for Burgers turbulence ($\ell_\text{c}^\text{B}=\text{Re}^{-2/3}L$). The inset shows a zoom of the dissipative range.}
\label{fig1}
\end{figure}

\begin{figure}
  \includegraphics[width=0.48\textwidth]{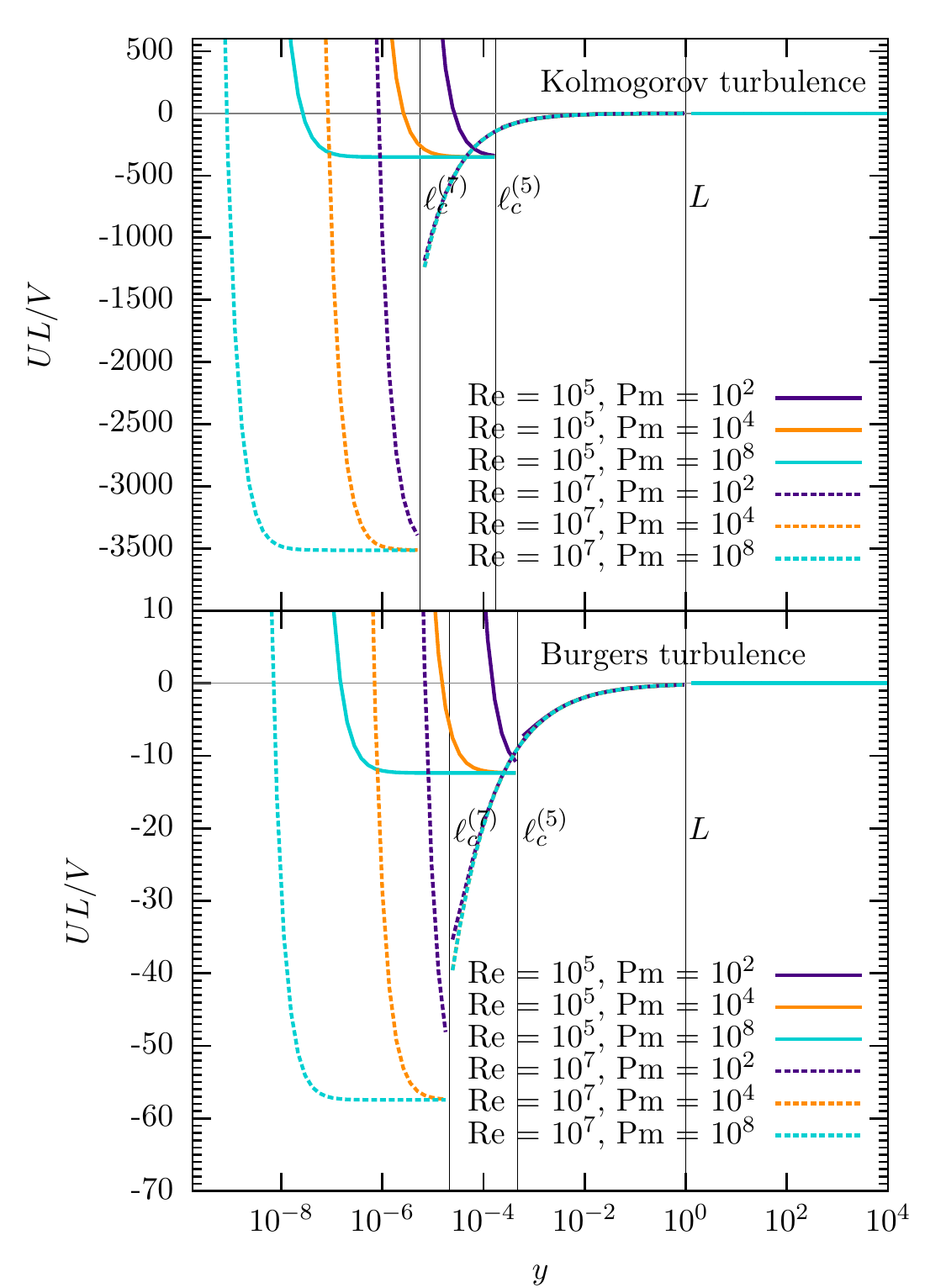}
  \caption{(Color online) Dependence of the potential on the dimensionless parameter $y\equiv r/L$ for Kolmogorov ($\vartheta=1/3$) and Burgers ($\vartheta=1/2$) turbulence. We choose two different Reynolds numbers $\text{Re}=10^5$ and $\text{Re}=10^{7}$, and different Prandtl numbers $\text{Pm}=10^2$, $\text{Pm}=10^4$ and $\text{Pm}=10^8$. The cutoff scale $\ell_\text{c}$ depends on the turbulence model and the Reynolds number. For Kolmogorov turbulence $\ell_\text{c}=\text{Re}^{-3/4}L$; for  Burgers turbulence $\ell_\text{c}=\text{Re}^{-2/3}L$. A Reynolds number $10^x$ is indicated in the cutoff scale as $\ell_\text{c}^\text{(x)}$.}
\label{fig2}
\end{figure}

We analyze the case of general types of turbulence, which can be described by the relation between the velocity $v(\ell)$ and the size $\ell$ of a turbulent fluctuation,
\begin{equation}
  v(\ell) \propto \ell^{\vartheta}.
\label{TurbPower}
\end{equation}
The power-law index $\vartheta$ varies for the different types. It attains its minimal value of $\vartheta=1/3$ for Kolmogorov theory \cite{Kolmogorov1941}, i.e., incompressible turbulence. For Burgers turbulence \cite{Burgers1948}, i.e., highly compressible turbulence, $\vartheta$ gets its maximal value of $1/2$ \cite{Schmidt2009}. \\
Motivated by the definition of the scale-dependent turbulent diffusion coefficient in the last section,
\begin{equation}
  \eta_\text{turb}(r) = T_\text{L}(0)- T_\text{L}(r),
\end{equation}
we construct a model for the longitudinal correlation function of the turbulent velocity field $T_\text{L}(r)$. The diffusion coefficient is calculated from the power law (\ref{TurbPower}) in the following way:
\begin{equation}
 \eta_\text{turb}(r) \propto v_\ell \ell \propto \ell^{\vartheta} \ell = \ell^{\vartheta + 1}.  
\end{equation}
So we assume the correlation function in the inertial range to be \cite{Vainshtein1982,Subramanian1997}
\begin{equation}
 T_\text{L}(r) = \frac{VL}{3} \left(1-\left(r/L\right)^{\vartheta+1}\right).
\end{equation}
The pre factor $VL$ fixes the units, which should be the same as for diffusivity. $V$ and $L$ are the velocity and the length scale of the largest eddies. On the diffusive scale the correlation function should be steadily continued and satisfy the condition that its derivative $T_{\text{L}}'(0)$ vanishes at $r=0$. This is accomplished, for example, for $T_{\text{L}} \propto r^2$. The exact form of $T_{\text{L}}$ in the diffusive range does not affect the results crucially \cite{Subramanian1997}. Furthermore, we expect no correlation on scales larger than the largest eddies, so $T_{\text{L}}$ should vanish there. \\
Taken all together, we can set up a general turbulence model for the longitudinal correlation function on the different length scales as follows:
\begin{equation}
  T_\text{L}(r) = \begin{cases} \frac{VL}{3}\left(1-\text{Re}^{(1-\vartheta)/(1+\vartheta)}\left(\frac{r}{L}\right)^{2}\right) 
                                                                                                                    & 0<r<\ell_\text{c} \\ 
  															\frac{VL}{3}\left(1-\left(\frac{r}{L}\right)^{\vartheta+1}\right) & \ell_\text{c}<r<L \\ 
  															0 & L<r, 
  							  \end{cases}
\label{TL}
\end{equation}
where $\ell_\text{c}=L~\text{Re}^{-1/(\vartheta+1)}$ denotes the cutoff scale of the turbulence and $L$ the length of the largest eddies. The hydrodynamic Reynolds number Re is defined as $(VL)/\nu$ with the typical velocity of the largest eddies $V$ and the viscosity of the gas $\nu$.\\
The transverse correlation functions $T_\text{N}$ for a divergence-free (i.e., Kolmogorov turbulence) and for an irrotational (i.e., Burgers turbulence) turbulent velocity field can be derived from the relations (\ref{TNdivfree}) and (\ref{TNrotfree}). Notice, however, that a turbulent velocity field that is divergence free or irrotational in the inertial range does not have to be this in the diffusive range. We find for the extreme cases in the inertial range ($\ell_\text{c}<r<L$)
\begin{eqnarray}
  T_\text{N}^{\text{K}}(r) & = & \frac{VL}{3}\left(1-\frac{5}{3}\left(\frac{r}{L}\right)^{4/3}\right), \label{TNKolmo} \\
  T_\text{N}^{\text{B}}(r) & = & \frac{VL}{3}\left(1-\frac{2}{5}\left(\frac{r}{L}\right)^{3/2}\right). 
\label{TNBurgers}
\end{eqnarray}
In order to find a general expression for $T_\text{N}$ we make the ansatz
\begin{equation}
  T_\text{N}(r) = \frac{VL}{3}\left(1-t(\vartheta)\left(\frac{r}{L}\right)^{\vartheta+1}\right), 
\end{equation}
where $t(\vartheta)=a-b\vartheta$.
With Eqs. (\ref{TNKolmo}) and (\ref{TNBurgers}) we find that $a=21/5$ and $b=38/5$. Furthermore, we find the small-scale transverse correlation (i.e., $0<r<\ell_\text{c} $) by steady continuation. So we end up with the following model for the transverse correlation function for the general slope of the turbulent velocity spectrum:
\begin{equation}
  T_\text{N}(r) = \begin{cases} \frac{VL}{3}\left(1-t(\vartheta)\text{Re}^{(1-\vartheta)/(1+\vartheta)}                                                                                                          \left(\frac{r}{L}\right)^{2}\right)     & 0<r<\ell_\text{c} \\ 
  															\frac{VL}{3}\left(1-t(\vartheta)\left(\frac{r}{L}\right)^{\vartheta+1}\right) 
  														                                                                         & \ell_\text{c}<r<L \\ 
  															0 & L<r, 
  							  \end{cases}
\label{TN}
\end{equation}
with $t(\vartheta)=(21-38\vartheta)/5$.\\
The longitudinal and transverse correlation functions depend on the dimensionless parameter $y \equiv r/L$ as shown in Fig.~\ref{fig1} for Kolmogorov and Burgers turbulence. We choose here a fixed hydrodynamical Reynolds number of $10^5$. In the inset of Fig.~\ref{fig1} we show a zoom into the dissipative range ($0<r<\ell_\text{c}$). Furthermore, in Fig.~\ref{fig2} we plot the potential of the Kazantsev equation, resulting from our model of $T_\text{L}$ for Kolmogorov and Burgers turbulence. We choose two different values for the Reynolds number Re and different magnetic Prandtl numbers Pm. The magnetic Prandtl number is defined as the ratio of the magnetic Reynolds number $\text{Rm}\equiv(VL)/\eta$ and the hydrodynamical Reynolds number Re. The potential at fixed Re and Pm is deeper in the small-scale range in the Kolmogorov case than in the Burgers case. For higher Re the potential gets deeper and the cutoff scale $\ell_\text{c}$ decreases. For higher Pm the potential in the small-scale range gets broader. 

\section{\label{sec: Results} Solution of the Kazantsev Equation}

In this section we use our model of the turbulent velocity correlation function (\ref{TL}) and (\ref{TN}) as the input for the Kazantsev theory. We solve the Kazantsev equation in order to obtain the characteristic properties of the small-scale dynamo. We use the WKB method, which gives a good approximate solution for large magnetic Prandtl number. In fact, in the limit of infinite magnetic Prandtl number the WKB approximation is an exact solution of the Kazantsev equation. For more details about the validity of this approximation, see the Appendix.

\subsection{Critical magnetic Reynolds number for small-scale dynamo action}

\begin{figure}[ht]
    \centering
    \includegraphics[width=0.48\textwidth]{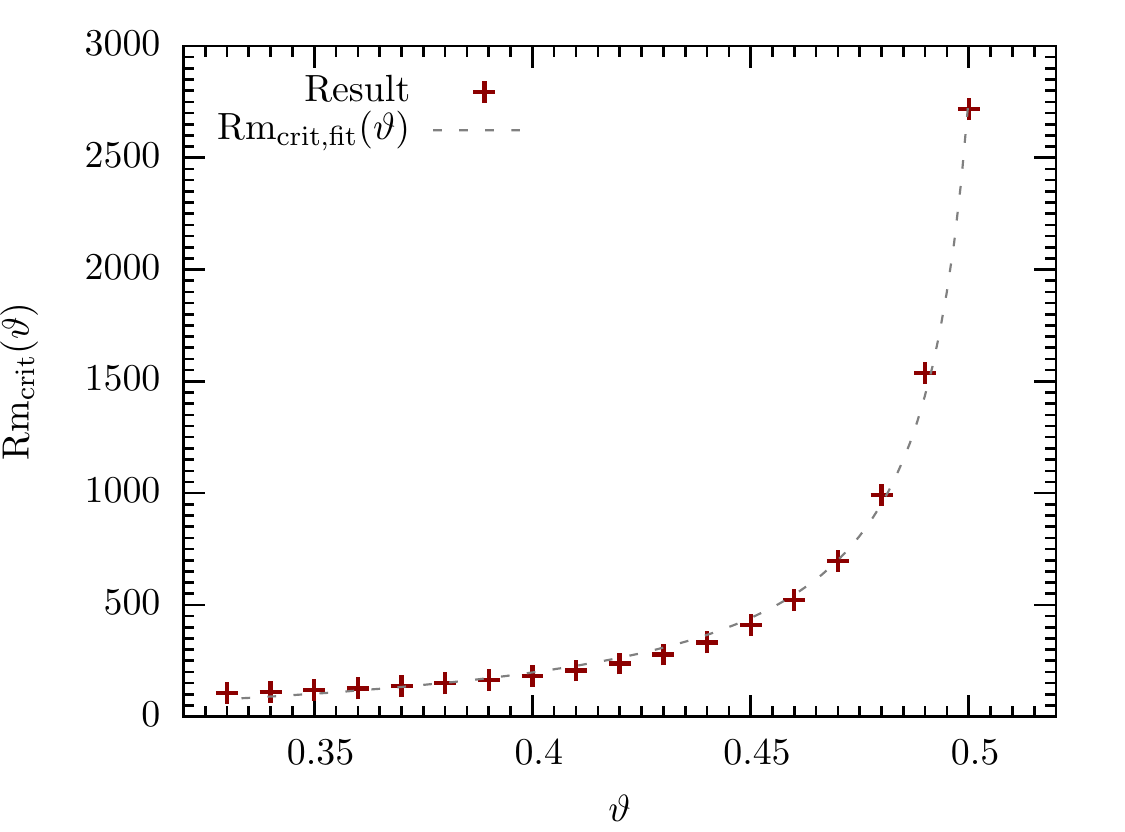}
    \caption{(Color online) Dependence of the critical magnetic Reynolds number $\text{Rm}_\text{crit}$ from the slope of the turbulent velocity spectrum                $\vartheta$. The dashed line is an empirical fit through our results.}
    \label{fig3}
\end{figure}

Intuitively, one expects that the high magnetic diffusivity for very low magnetic Reynolds numbers prevents amplification of the magnetic field. Even higher diffusivity eventually results in a net decrease of the field strength. In this section we calculate the critical magnetic Reynolds number $\text{Rm}_\text{crit}$ for small-scale dynamo action. To accomplish this we set the growth in our equations at zero. \\
It should be noted that we use the inertial range ($\ell_\text{c}<r<L$) for determining $\text{Rm}_\text{crit}$ as the potential in this range is always negative and for that we have a positive growth rate (see Fig.~\ref{fig2}). In this range with our turbulence spectrum and $\Gamma=0$ we get for the $p$-function (\ref{p})
\begin{eqnarray}
  p(y) & = &  \frac{-9/4-a(\vartheta)\text{Rm}_{\text{crit}}y^{\vartheta+1} + b(\vartheta)\text{Rm}_{\text{crit}}^2 y^{2(\vartheta+1)}}
             {\left(1 + \frac{1}{3}\text{Rm}_{\text{crit}}y^{\vartheta+1}\right)^2} \nonumber \\
       &   &  
\end{eqnarray}
with $a(\vartheta) \equiv 5/6-(79/30)~\vartheta+(157/30)~\vartheta^2$ and $b(\vartheta) \equiv (14/15)~\vartheta-(103/60)~\vartheta^2$.\\
Now we can evaluate the eigenvalue condition (\ref{eigenvalue}) for this $p(y)$ in the ``ground state" $n=0$:
\begin{eqnarray}
  \int_{y_{0_1}}^{y_{0_2}}\sqrt{p(y)}\frac{\text{d}y}{y} = \frac{\pi}{2},
\label{EigenvalueCritMRN}
\end{eqnarray}
in which the additional $y$ comes from the substitution $y=r/L=e^x/L$. The limits of the integral are the roots of $p(y)$. There is only one real and positive root of $p(y)$, which we label $y_1$. For the upper limit we have to realize that the potential (\ref{GeneralPotential}) changes for $y>1$ to $2\eta/(yL)^2$, which is clearly always positive. Furthermore, also the diffusion coefficient $\kappa_\text{diff}=\eta + T_\text{L}(0)>0$ for $y>1$. With $U$ and $\kappa_\text{diff}$ being positive $p(y)$ is negative in this range, which means that $p(y)$ needs to go through zero during this transition. So we have our second root at roughly $r\approx L$ and $y_2=1$. \\
We can solve Eq.~(\ref{EigenvalueCritMRN}) numerically for the critical magnetic Reynolds number $\text{Rm}_\text{crit}$ if we put in a fixed value of $\vartheta$. Recall that $\vartheta$ was defined in the inertial range of the turbulence via the relation $v(\ell)\propto \ell^\vartheta$. Results for common models in the literature can be found in Table \ref{ResultsTable}. In Fig.~\ref{fig3} we show how the critical magnetic Reynolds number depends on $\vartheta$. Here one can see that the critical magnetic Reynolds number increases rapidly as $\vartheta$ gets closer to its maximum value of $1/2$. An empirical fit $\text{Rm}_\text{crit,fit}(\vartheta)$ through these data in the range $0.33 < \vartheta < 0.5$ is
\begin{equation}
  \text{Rm}_\text{crit,fit}(\vartheta) = 88\left(\text{tan}(2.68\vartheta+0.2)-1\right).
\end{equation}
Furthermore, we collect the results for common turbulence models in the literature in Table \ref{ResultsTable}. \\
We find that the small-scale dynamo is more easily excited in the case of a purely rotational turbulent velocity field, i.e., for Kolmogorov turbulence, where we find $\text{Rm}_\text{crit}\approx110$. The critical magnetic Reynolds number for a turbulent field with a vanishing rotational component, i.e., Burgers turbulence, is roughly $2700$.\\
The results are discussed in Sec.~\ref{sec: Discussion}.

\subsection{Growth rate of the small-scale magnetic field}

\subsubsection{Growth Rate in the Limit $\text{Pm}\rightarrow\infty$}
In this section we derive a general analytical solution for the growth rate $\Gamma$ for an arbitrary slope of the turbulent velocity spectrum, in the limit of infinite Prandtl number. \\
As the potential has its minimum in the small-scale range, i.e., the dissipative range of the turbulence (see Fig.~\ref{fig2}), the growth rate, which is the eigenvalue of the Kazantsev equation, takes its maximal value there. So we expect the fastest growing mode to be in the small-scale range.\\
In order to have scale-independent equations, we introduce the substitution
\begin{equation}
	z \equiv \left(\frac{V\sqrt{\text{Re}}}{3L\eta}\right)^{1/2}r = \left(\frac{\text{Re}^{3/2}\text{Pm}}{3}\right)^{1/2}y,
\label{z}
\end{equation}
where the magnetic Prandtl number is Pm=Rm/Re.\\
The $p$ function in $z$ space [see (\ref{p}) and (\ref{z})] for the general turbulence spectrum (\ref{TL}) in the dissipative range is 
\begin{eqnarray}
  p(z) & = & \frac{A_0 z^4 - B_0 z^2 - 45 \text{Re}^{(3+7\vartheta)/(2+2\vartheta)}}{20 \text{Re}^{1/2}\left(\text{Re}^{(1+3\vartheta)/(2+2\vartheta)} +                     \text{Re}^{1/(1+\vartheta)}z^2\right)^2}       \nonumber \\
       &   & 
\label{pModel}             
\end{eqnarray}
with the definitions
\begin{eqnarray}
  A_0 & = & \text{Re}^{(5+\vartheta)/(2+2\vartheta)}\left(163-304\vartheta\right)-\frac{20}{3} \text{Re}^{5/2} \bar{\Gamma}, \\
  B_0 & = & \left(304\vartheta-98\right)\text{Re}^2+\frac{20}{3}\text{Re}^{(2+8\vartheta)/(1+\vartheta)}\bar{\Gamma},
\end{eqnarray}
and the normalized growth rate
\begin{equation}
  \bar{\Gamma} \equiv \frac{L}{V}\Gamma.
\label{normGamma}
\end{equation}
In the limit of large Prandtl number $z$ is large, too, and we can neglect the constant terms. We obtain
\begin{equation}
  p(z) =  \frac{\text{Re}^{-(5+\vartheta)/(2+2\vartheta)}}{20} \frac{A_0z^2-B_0}{z^2}.
\end{equation}
The one real and positive root of this function is $z_1=\sqrt{B_0/A_0}$. At the cutoff scale of the turbulence the $p$ function changes its sign. We take this as our second root and so have $z_2=\sqrt{\text{Pm}/3}~\text{Re}^{(3\vartheta-3)/(4\vartheta+4)}$. So we get for the general eigenvalue condition
\begin{equation}
  \frac{\text{Re}^{-(5+\vartheta)/(4+4\vartheta)}}{2\sqrt{5}}\int_{z_1}^{z_2}\sqrt{\frac{A_0z^2 - B_0}{z^4}}\text{d}z = \frac{\pi}{2},
\end{equation}
resulting in the analytical solution of the integral
\begin{eqnarray}
  \frac{\text{Re}^{-(5+\vartheta)/(4+4\vartheta)}}{2\sqrt{5}z}\left[\sqrt{A_0}\text{ln}\left(2\left(\sqrt{A_0}z+ \right.\right.\right. \nonumber \\   \left.\left.\left.\left.\sqrt{A_0z^2-B_0}\right)\right) - \sqrt{A_0z^2-B_0}\right]\right|^{z_2}_{z_1} = \frac{\pi}{2}.
\end{eqnarray}
For $z_2\gg1$ this becomes
\begin{eqnarray}
  \frac{\text{Re}^{-(5+\vartheta)/(4+4\vartheta)}}{2\sqrt{5}}\sqrt{A_0}\left[1-\text{ln}\left(4\sqrt{A_0}z_2\right)\right. \nonumber \\ 
  \left. + \frac{1}{2} \text{ln}\left(4B_0\right)\right] =  \frac{\pi}{2}.  
\end{eqnarray}
A zero-order iterative solution for $\bar{\Gamma}$ gives us
\begin{eqnarray}
  \bar{\Gamma} & = & \frac{163-304\vartheta}{60}\text{Re}^{(1-\vartheta)/(1+\vartheta)} - \\ 
               &   & \left(\frac{\pi\sqrt{5}}{\text{Re}^{-(5+\vartheta)/(4+4\vartheta)}\left[1-\text{ln}\left(4\sqrt{A_0}z_2\right) + 1/2~                       \text{ln}\left(4B_0\right)\right]}\right)^2, \nonumber
\end{eqnarray}
which becomes for large Prandtl number
\begin{equation}
  \bar{\Gamma} = \frac{163-304\vartheta}{60}\text{Re}^{(1-\vartheta)/(1+\vartheta)}.
\end{equation}
As a result we get for the absolute growth rate $\Gamma$ for a general slope of the turbulent velocity spectrum
\begin{eqnarray}
  \Gamma =  \frac{(163-304\vartheta)~V}{60~L}\text{Re}^{(1-\vartheta)/(1+\vartheta)}
\label{Gamma}
\end{eqnarray}
in the limit $\text{Pm}\rightarrow\infty$.
\begin{figure}[ht]
    \centering
    \includegraphics[width=0.48\textwidth]{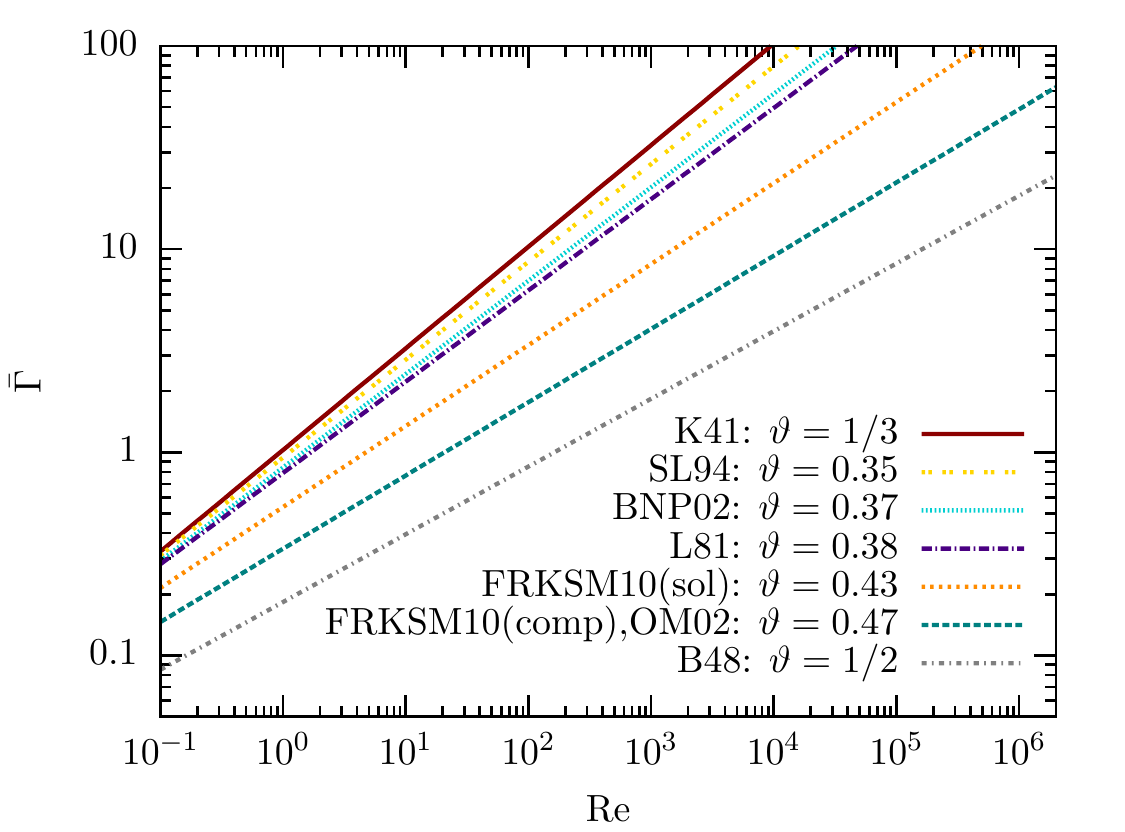}
    \caption{(Color online) The normalized growth rate of the small-scale dynamo in the limit of infinite magnetic Prandtl number, depending on the Reynolds number Re. For the slopes of the turbulent velocity spectrum $\vartheta$ we choose common values from the literature: K41 \cite{Kolmogorov1941}, SL94 \cite{SheLeveque1994}, BNP02 \cite{Boldyrev2002}, L81 \cite{Larson1981}, FRKSM10 \cite{Federrath2010} (sol: solenoidal forcing; comp: compressive forcing), OM02 \cite{OssenkopfMacLow2002} and B48 \cite{Burgers1948}.}
    \label{fig4}
\end{figure}
In Fig.~\ref{fig4} we show the dependency of the normalized growth rate $\bar{\Gamma}$ on the Reynolds number for different types of turbulence. One extreme case is incompressible turbulence, i.e., Kolmogorov turbulence, with $\bar{\Gamma}\propto \text{Re}^{1/2}$. In the other extreme case, highly compressible turbulence, i.e., Burgers turbulence, the growth rate increases only as $\text{Re}^{1/3}$. Altogether we find that the growth rate increases faster with the Reynolds number when the compressibility is lower.

\subsubsection{Growth rate as a function of the Prandtl number}
\begin{figure}[ht!]
    \centering
    \includegraphics[width=0.48\textwidth]{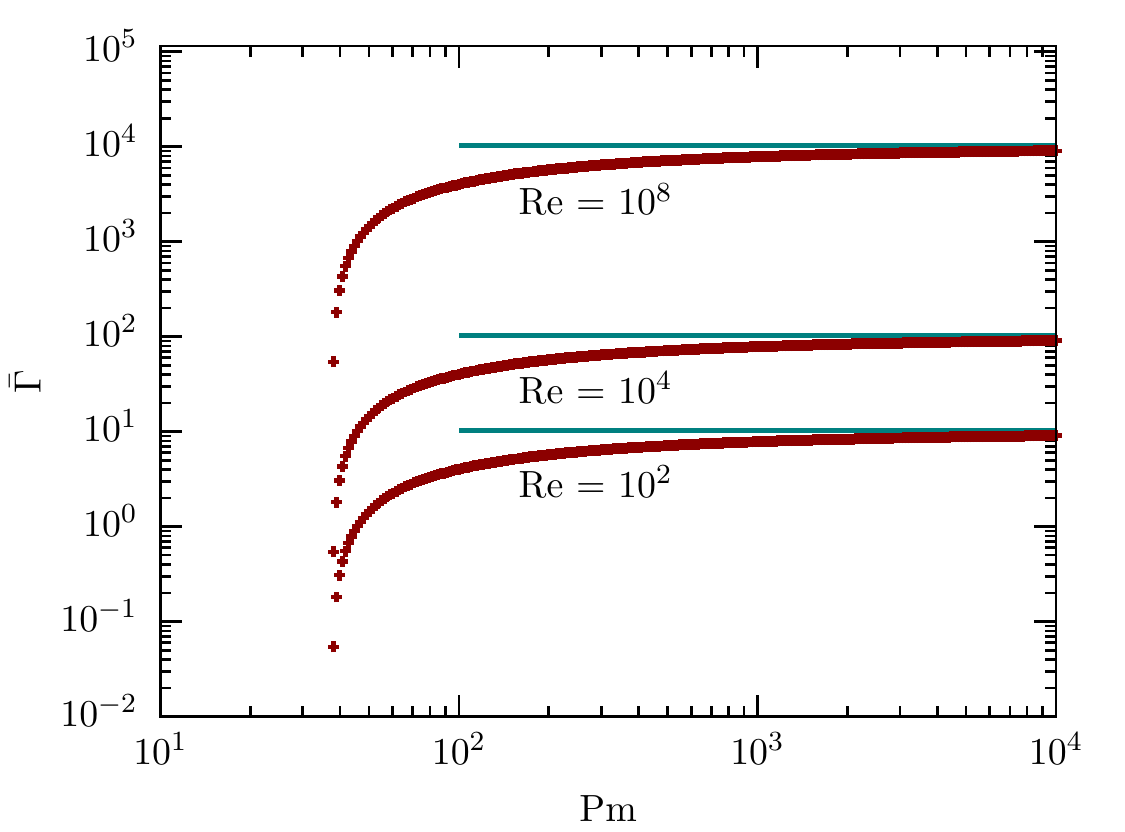}
    \caption{(Color online) Dependence of the normalized growth rate of the small-scale dynamo on the magnetic Prandtl number Pm for Kolmogorov turbulence. We choose different values of the Reynolds number. Notice that in the limit $\text{Pm}\rightarrow\infty$ the normalized growth rates are $\bar{\Gamma}=10.28$ for $\text{Re}=10^{2}$, $\bar{\Gamma}=102.78$ for $\text{Re}=10^{4}$, and $\bar{\Gamma}=10277.78$ for $\text{Re}=10^{8}$. These limits are indicated in the plot as horizontal lines.}
    \label{fig5}
\end{figure}

\begin{figure}[ht!]
    \centering
    \includegraphics[width=0.48\textwidth]{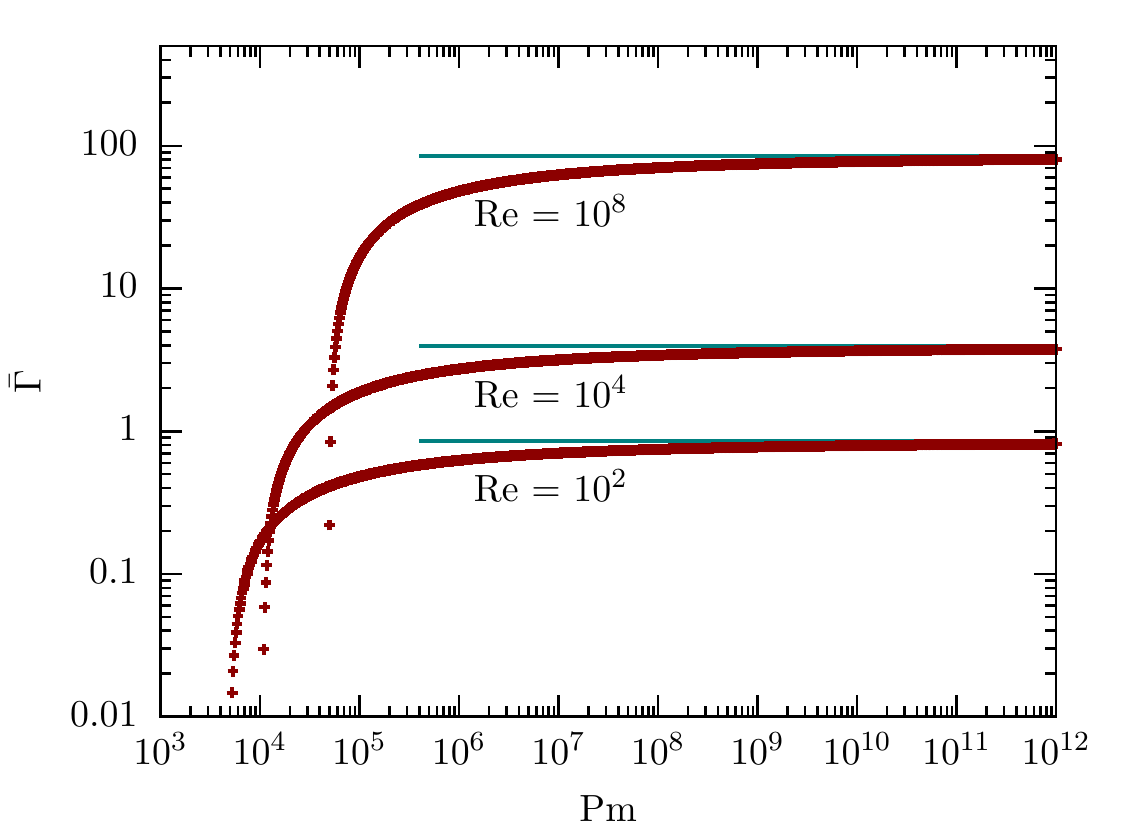}
    \caption{(Color online) Dependence of the normalized growth rate of the small-scale dynamo on the magnetic Prandtl number Pm for Burgers turbulence. We choose different values of the Reynolds number. Notice that in the limit $\text{Pm}\rightarrow\infty$ the normalized growth rates are $\bar{\Gamma}=0.85$  for $\text{Re}=10^{2}$, $\bar{\Gamma}=3.95$ for $\text{Re}=10^{4}$, and $\bar{\Gamma}=85.1$ for $\text{Re}=10^{8}$. These limits are indicated in the plot as horizontal lines.}
    \label{fig6}
\end{figure}

In this section we discard the assumption of infinite Prandtl number. In this case we have to solve the full equation resulting from the WKB method (\ref{eigenvalue}),
\begin{equation}
  \int\frac{\sqrt{p(z)}}{z}\text{d}z = \frac{\pi}{2},
\end{equation}
with $p(z)$ from (\ref{pModel}). There is no analytical solution of this integral equation.\\
The numerical results of the normalized growth rate are shown in Fig.~\ref{fig5} for Kolmogorov turbulence and in Fig.~\ref{fig6} for Burgers turbulence. We plot the normalized growth rate depending on the Prandtl number for different values of the Reynolds number.

\section{\label{sec: Discussion} Discussion}

The main results of this work are the critical magnetic Reynolds number $\text{Rm}_\text{crit}$ and the growth rate $\Gamma$ of the small-scale dynamo given by Fig.~\ref{fig3} and (\ref{Gamma}). In Table \ref{ResultsTable}, we collect our results for typical turbulence models from the literature, defined via the relation $v(\ell)\propto\ell^\vartheta$. The two extrema of turbulence are Kolmogorov turbulence \cite{Kolmogorov1941}, i.e., incompressible turbulence, with $\vartheta=1/3$ and Burgers turbulence with $\vartheta=1/2$, which describes highly compressible turbulence with vanishing rotational component \cite{Burgers1948}. Between those extreme values we choose values of $\vartheta$ from observations of molecular clouds, like $\vartheta\approx0.38$ from Larson \cite{Larson1981} and $\vartheta\approx0.47$ from Ossenkopf and Mac Low \cite{OssenkopfMacLow2002}. Furthermore, we give $\vartheta\approx0.35$ as an example for a theoretical model of intermittency \cite{SheLeveque1994}. Numerical experiments give $\vartheta\approx0.37$ for driven supersonic magnetohydrodynamical (MHD) turbulence \cite{Boldyrev2002} and $\vartheta\approx0.43$ and $\vartheta\approx0.47$ for solenoidal and compressive forcing of the turbulence \cite{Federrath2010}. Notice, however that the mean values of $\vartheta$ from observations and simulations have a typical uncertainty of 10\%. 

\begin{table*}  
  \begin{ruledtabular}
    \begin{tabular}{lccc}
      \parbox[0pt][2.5em][c]{0cm}{}  Model and reference                 & $\vartheta$     &  $\text{Rm}_\text{crit}$     & $\bar{\Gamma}$                                                                                                                             ($\text{Pm}\rightarrow\infty$)\\
      \hline
      \tabhe Kolmogorov \cite{Kolmogorov1941}			    								&  $1/3$          &  $\approx 107$     &  $\frac{37}{36}~\text{Re}^{1/2}$ \\
      \tabhe Intermittency of Kolmogorov turbulence (She and Leveque \cite{SheLeveque1994})  &  $0.35$ & $\approx 118$ & $0.94~\text{Re}^{0.48}$\\
      \tabhe Driven supersonic MHD-turbulence (Boldyrev\textit{ et al.\ }\cite{Boldyrev2002})&  $0.37$  &  $\approx 137$    &  $0.84~\text{Re}^{0.46}$   \\
      \tabhe Observation in molecular clouds (Larson \cite{Larson1981})  &  $0.38$  &  $\approx 149$     &  $0.79~\text{Re}^{0.45}$     \\
      \tabhe Solenoidal forcing of the turbulence (Federrath\textit{ et al.\ }\cite{Federrath2010}) &  $0.43$ &  $\approx 227$  &                                                                                                                                  $0.54~\text{Re}^{0.40}$  \\
      \tabhe Compressive forcing of the turbulence (Federrath\textit{ et al.\ }\cite{Federrath2010}), &  $0.47$   &  $\approx 697$  &                                                                                                                              $0.34~\text{Re}^{0.36}$  \\                     
             Observations in molecular clouds (Ossenkopf and Mac Low \cite{OssenkopfMacLow2002})    &       &       &             \\ 
      \tabhe Burgers \cite{Burgers1948}                             &  $1/2$     &  $\approx 2718$   &  $\frac{11}{60}~\text{Re}^{1/3}$ \\
    \end{tabular}
  \end{ruledtabular}\caption{The critical magnetic Reynolds number $\text{Rm}_\text{crit}$ and the normalized growth rate of the small-scale dynamo $\bar{\Gamma}$ in the limit of infinite magnetic Prandtl number. We show our results for different types of turbulence, which are characterized by the exponent $\vartheta$ of the slope of the turbulent velocity spectrum, $v(\ell)\propto\ell^\vartheta$. The extreme values of $\vartheta$ are $1/3$ for Kolmogorov turbulence and $1/2$ for Burgers turbulence.}
  \label{ResultsTable}
\end{table*}

\subsection{\label{DisRmcrit} Critical magnetic Reynolds number}

We have evaluated Eq.~(\ref{EigenvalueCritMRN}) for different slopes of the turbulent velocity spectrum $\vartheta$. The results for common types of turbulence from the literature are listed in Table \ref{ResultsTable} and pictured in Fig.~\ref{fig3}. \\
From these results we see that for all types of turbulence a high magnetic Reynolds number needs to be exceeded for small-scale dynamo action. In astrophysical objects we often find very high magnetic Reynolds numbers (see the compilation in \cite{ChildressGilbert1995}). The core of Jupiter, for example, has $\text{Rm}\approx 10^6$, the solar convection zone has $\text{Rm}\approx 10^8$, and the solar corona already $\text{Rm}\approx 10^{12}$. In the interstellar medium we find $\text{Rm}\approx 10^{17}$ and in a typical galaxy about $10^{19}$. Consequently, the critical magnetic Reynolds number is exceeded by far in nature, and we expect that a small-scale dynamo is excited in typical astrophysical objects.\\
There are different ways to obtain approximate solutions of the Kazantsev equation. In addition to the WKB method there is also an asymptotic solution, which uses the separation of scales \cite{ZeldovichRuzmaikinSokoloff1990}. The potential $U(r)$ and the mass $m(r)$ are estimated on different scales, such that a solution of the Kazantsev equation can be found. Rogachevskii and Kleeorin \cite{RogachevskiiKleerin1997} used this method to determine the critical magnetic Reynolds number for different compressibilities of turbulence. They found that, in the limit of small magnetic Prandtl number \text{Rm} needs to be larger than roughly 400 for excitation of a small-scale dynamo in the case of Kolmogorov turbulence. For a larger compressibility, i.e., toward Burgers turbulence, they found that $\text{Rm}_\text{crit}$ increases sharply. We see the same trend of increasing $\text{Rm}_\text{crit}$ for higher compressibility. However, our result for Kolomogorov turbulence differs by a factor of roughly 4, which might be caused by the fact that we analyze the limit of large magnetic Prandtl number.\\
Recent high-resolution numerical studies confirm the existence of a critical magnetic Reynolds number for small-scale dynamo action. Haugen\textit{ et al.\ }\cite{Haugen2004} found $\text{Rm}_\text{crit}\approx 35$ for subsonic turbulence and $\text{Rm}_\text{crit}\approx 70$ for supersonic turbulence at a magnetic Prandtl number of about unity. In numerical simulations, the magnetic Reynolds number can be estimated by $\text{Rm}\approx(\lambda/\ell_\text{c})^{\vartheta+1}$, where $\lambda$ is the typical size of turbulent structures and $\ell_\text{c}$ is the cutoff scale of the turbulence. The latter can be estimated by $\ell_\text{c}\approx0.5~\Delta x$ with $\Delta x$ the minimal resolved size in a simulation \cite{Benzi2008}. In resolution studies, Sur\textit{ et al.\ }\cite{Sur2010} and Federrath\textit{ et al.\ }\cite{Federrath2011.1} found that the typical length of a turbulent fluctuation needs to be resolved with at least $30$ grid cells in magnetohydrodynamic simulations of a self-gravitating gas. Only then is the magnetic field amplified exponentially, which is explained by the action of the small-scale dynamo. \\
For a physical interpretation of this result it is useful to take the \textit{stretch-twist-fold-dynamo} as a toy-model of the turbulent dynamo. In this picture we think of a magnetic flux rope that gets stretched due to turbulent motions. This motion decreases the cross section of the flux rope. If we assume that the magnetic flux is frozen in the gas, then the magnetic field increases during this process, because the magnetic flux is a conserved quantity \cite{ChildressGilbert1995,BrandenburgSubramanian2005}. This process works best in a purely rotational turbulent velocity field. Therefore, we expect the dynamo to be more easily excited in Kolmogorov turbulence. In order to see this process in simulations, one needs to resolve the stretching, twisting, and folding of the field lines, which explains the required high resolution.\\
The determination of the critical magnetic Reynolds number is, moreover, the first step to understanding the saturation of the small-scale dynamo. For if the magnetic field in a system increases, back reactions from the gas become more important. Then processes like ambipolar diffusion can change the properties of the gas and the magnetic Reynolds number can decrease. If the magnetic Reynolds number becomes smaller than the critical magnetic Reynolds number, the magnetic field stops growing and the small-scale dynamo is saturated.

\subsection{Growth rate}

In Fig.~\ref{fig4} as well as in Table \ref{ResultsTable}, we present our results for the growth rate of the small-scale dynamo in the limit of infinite magnetic Prandtl number. Our results show that the growth rate is proportional to the velocity $V$ of the largest eddy divided by its length $L$. The ratio $V/L$ is the reciprovcal of the turnover time of an eddy. Thus, the growth rate increases with decreasing turnover time, and the smallest modes grow at the highest rate. This is expected, because smaller turnover times lead to a faster tangling of the magnetic field lines.\\ 
Furthermore, the growth rate increases with increasing hydrodynamical Reynolds number for all types of turbulence, characterized by $v(\ell)\propto \ell^{\vartheta}$. In order to achieve the same growth rate for Kolmogorov and Burgers turbulence we have to provide a larger Reynolds number in the latter case. Assuming a fixed Reynolds numbers in both extreme cases, $\text{Re}^\text{K}$ and $\text{Re}^\text{B}$, the growth rates of the two different turbulence types are the same for $\text{Re}^\text{K} \approx 0.18(\text{Re}^\text{B})^{3/2}$. This fact can again be explained with the stretch-twist-fold model (see Sec.~ \ref{DisRmcrit}). We need solenoidal modes, i.e., divergence-free modes, of the turbulence for this process \cite{Federrath2011.2}, which explains why incompressible turbulence amplifies the magnetic field more effectively. \\
With the asymptotic solution of the Kazantsev equation, Rogachevskii and Kleeorin \cite{RogachevskiiKleerin1997} found in the limit of small magnetic Prandtl number $\Gamma \propto \text{ln}(\text{Rm}/\text{Rm}_\text{crit})$. The constant of proportionality depends on the amount of compressibility. In a later work Kleeorin and Rogachevskii \cite{KleeorinRogachevskii2011} found that this logarithmic scaling of the growth rate is valid only in the vicinity of the threshold of small-scale dynamo excitation. For magnetic Reynolds numbers much larger than $\text{Rm}_\text{crit}$, they found in the limit of small magnetic Prandtl number $\Gamma \propto \text{Rm}^{1/2}$ for Kolmogorov turbulence. As for a constant magnetic Prandtl number $\text{Rm} \propto \text{Re}$, this agrees with our result.\\
There are recent high-resolution numerical simulations that model the turbulent dynamo. The two limiting ways of driving the turbulence are solenoidal, i.e., divergence-free, forcing and compressive, i.e., rotation-free, forcing. These simulations show, in agreement with our study, that solenoidally driven turbulence leads to larger growth rates of the small-scale dynamo. Waagan\textit{ et al.\ }\cite{Waagan2011} found, using a Reynolds number of about $1500$ \footnote{Waagan\textit{ et al.\ }\cite{Waagan2011} gave a magnetic Reynolds number of about $200$. However, these ideal MHD simulations were later calibrated with resistive non-ideal MHD-simulations in Ref.~\cite{Federrath2011.2}, showing that the Reynolds number is about $1500$.}, and a magnetic Prandtl number of about $1$, for totally solenoidal forcing of the turbulence $\bar{\Gamma}^\text{sol}=0.60$ and for totally compressive forcing  $\bar{\Gamma}^\text{comp}=0.28$. These values of the growth rate are about a factor of $17$ lower than those from our model (with $\text{Re}=1500$), $\bar{\Gamma}^\text{sol} = \bar{\Gamma}^{\vartheta=0.43} \approx 10.07$ and $\bar{\Gamma}^\text{comp} = \bar{\Gamma}^{\vartheta = 0.47} \approx 4.73$. This can be explained by the fact that the simulations had a very low magnetic Prandtl number of about $1$. However, our result for the growth rate in Table \ref{ResultsTable} was derived with the assumption of infinite Prandtl number. We have also explored the range of lower Prandtl numbers. The result is presented in Fig.~\ref{fig5} for Kolmogorov turbulence and in Fig.~\ref{fig6} for Burgers turbulence. But with our model, we can make no predictions for Prandtl numbers around unity, because in this range the WKB approximation is no longer applicable (see the Appendix). However, the trend is that the growth rate decreases for lower Prandtl numbers and this can explain the lower growth rates from the simulations. Yet the ratio of the growth rate of turbulence driven by solenoidal and compressive forcing is in both cases about 2 (our model: $\bar{\Gamma}^\text{sol}/ \bar{\Gamma}^\text{comp} \approx 2.1$; Waagan\textit{ et al.\ }\cite{Waagan2011}: $\bar{\Gamma}^\text{sol}/ \bar{\Gamma}^\text{comp} \approx 2.1$), which shows that incompressible turbulence is more efficient in amplifying a magnetic field via the small-scale dynamo. \\
Furthermore, Federrath\textit{ et al.\ }\cite{Federrath2011.2} have presented a study of the Mach number dependency of the growth rate of the small-scale dynamo, where they compared solenoidal with compressive forcing of the turbulence. They found that, for low Mach numbers, the ratio of the growth rate of turbulence driven by solenoidal and compressive forcing is about $30$ (for Mach number $M=0.1$: $\bar{\Gamma}^\text{sol} \approx 1.2$, $\bar{\Gamma}^\text{comp} \approx 0.04$). However, for higher Mach numbers their calculations also result in a ratio of $\bar{\Gamma}^\text{sol} / \bar{\Gamma}^\text{comp} \approx 2$ (for Mach number $M=10$: $\bar{\Gamma}^\text{sol} \approx 0.7$, $\bar{\Gamma}^\text{comp} \approx 0.3$), which is in agreement with our results. The lower growth rates in the simulation again may come from low Prandtl numbers in the simulations, of the order of unity. However, a great uncertainty is the Reynolds number in numerical simulations, which is only a crude estimate.

\section{\label{sec: Con&Out} Summary and Conclusions}

We have presented an analytical treatment of the small-scale dynamo, using the Kazantsev theory. For that we have modeled the correlation function of the turbulent velocity field, depending on the slope of the turbulent velocity spectrum, $\vartheta$, in $v(\ell)\propto\ell^\vartheta$. With this model, we solved the Kazantsev equation in the WKB approximation and tested the validity of this approximation. We determined the critical magnetic Reynolds number for the small-scale dynamo and its growth rate in the case of infinite and finite magnetic Prandtl numbers.\\ 
The main results of our work are as follows:
\begin{itemize}
	\item The critical magnetic Reynolds number $\text{Rm}_\text{crit}$ for the small-scale dynamo increases as the exponent $\vartheta$ increases          (see Fig.~\ref{fig3}). For Kolmogorov turbulence $\text{Rm}_\text{crit}^\text{K}\approx110$ and for Burgers turbulence                           $\text{Rm}_\text{crit}^\text{B}\approx2700$.
	\item The growth rate of the magnetic field energy in the limit of infinite magnetic Prandtl number is  
	      \begin{equation}
	         \Gamma = \frac{(163-304\vartheta)~V}{60~L}~\text{Re}^{(1-\vartheta)/(1+\vartheta)}
	      \end{equation}
	      (see also Figure \ref{fig4}).
	\item For decreasing magnetic Prandtl number the growth rate decreases. The details of this drop depend on the type of turbulence (see           Figs.~ \ref{fig5} and \ref{fig6}).
	\item A validity test shows that the WKB-approximation gives exact solutions in the limit of infinite magnetic Prandtl number. The              approximation breaks down at a Prandtl number of around unity (see Figs.~\ref{fig7} and \ref{fig8}).
\end{itemize}
With these results we are able to make predictions about the first magnetic fields in the Universe. As the hydrodynamical Reynolds numbers are typically very high in astrophysical objects, the growth rates of the small-scale dynamo are very high, too. We thus expect that the time until saturation is shorter than the collapse time of a halo \cite{Federrath2011.1}. For that reason we might already have high magnetic field strengths even before the formation of the first stars, the first galaxies, and the first galaxy clusters.\\
Turbulence and magnetic fields are key ingredients of current star formation theory \cite{MacLowKlessen2004,McKeeOstriker2007}. Magnetic fields drive jets and outflows from young stars. Stellar winds and supernova explosions, which end the lives of massive stars, enrich the interstellar medium with heavy elements forged in the stellar interior. These processes are crucial for the chemical composition of the Universe, determining cooling and heating processes in the gas. This, in turn, is very important for the formation of the next generation of stars. The momentum from jets and outflows around accreting protostars may disperse some of the envelope material that otherwise would fall onto the central star. Thus, they are important ingredients for our understanding of the physical origin of the observed distribution of stellar masses \cite{Chabrier2003}.

\begin{acknowledgments}
J.S.~acknowledges financial support by the {\em Deutsche Forschungsgemeinschaft} (DFG) in the {\em Schwerpunktprogramm} SPP 1573 ``Physics of the Interstellar Medium" under Grant No.~KL 1358/14-1. D.R.G.S. acknowledges funding through the SPP 1573 (Project No. SCHL~1964/1-1) and the SFB 963 ``Astrophysical Flow Instabilities and Turbulence''. C.F.~acknowledges funding from the Australian Research Council (Grant No.~DP110102191) and from the European Research Council (FP7 Grant Agreement No.~247060). C.F.,~R.B.,~and R.S.K.~acknowledge subsidies from the Baden-W\"urttemberg-Stiftung under Research Contract No.~P-LS-SPII/18 and from the German Bundesministerium f\"ur Bildung und Forschung via the ASTRONET project STAR FORMAT (Grant No.~05A09VHA). R.S.K also thanks the DFG for financial support via Grants No.~KL1358/10 and No.~KL1358/11, as well as via the SFB 881 ``The Milky Way System". R.B. acknowledges funding by the Emmy-Noether Grant No.~(DFG) BA~3706 and the DFG via the Grands No.~BA 3706/1-1 and No.~BA 3706/3-1.
\end{acknowledgments}

\appendix

\section*{\label{sec. Validity} APPENDIX: Validity of the WKB Approximation} 

The WKB method is only an approximate solution of the Kazantsev equation. We have derived condition (\ref{CheckApprox}), $\left|f\right|\ll 1$, for which the WKB-method is valid in order to find solutions. In $z$ space, $f$ reads
\begin{equation}
  f(z) \equiv \frac{z^2p''(z)+2zp'(z)}{4p(z)^2} - \frac{3}{16}\frac{[z p'(z)]^2}{p(z)^3}.
\label{f}
\end{equation}
However, we have seen that the magnetic field is amplified most strongly on the scale $\ell_\text{c}(z)=\sqrt{\text{Pm}/3}$, as here the potential $U$ has its minimum. So we analyze $f(z,\Gamma)$ on this scale and get a dependency on the Prandtl number Pm. Hence we label $ f(\ell_\text{c},\Gamma) \equiv f(\text{Pm},\Gamma)$.\\
One can show that $f(\text{Pm},\Gamma)$ vanishes in the limit of large Prandtl number for all $\Gamma$ and all turbulence types,
\begin{equation}
  \lim\limits_{\text{Pm} \rightarrow \infty} f(\text{Pm},\Gamma) = 0.
\end{equation}
This means that the WKB method is very good in the limit of large magnetic Prandtl number. \\

\subsection{Validity of the WKB approximation for Kolmogorov turbulence}

\begin{figure}
  \centering
  \includegraphics[width=0.48\textwidth]{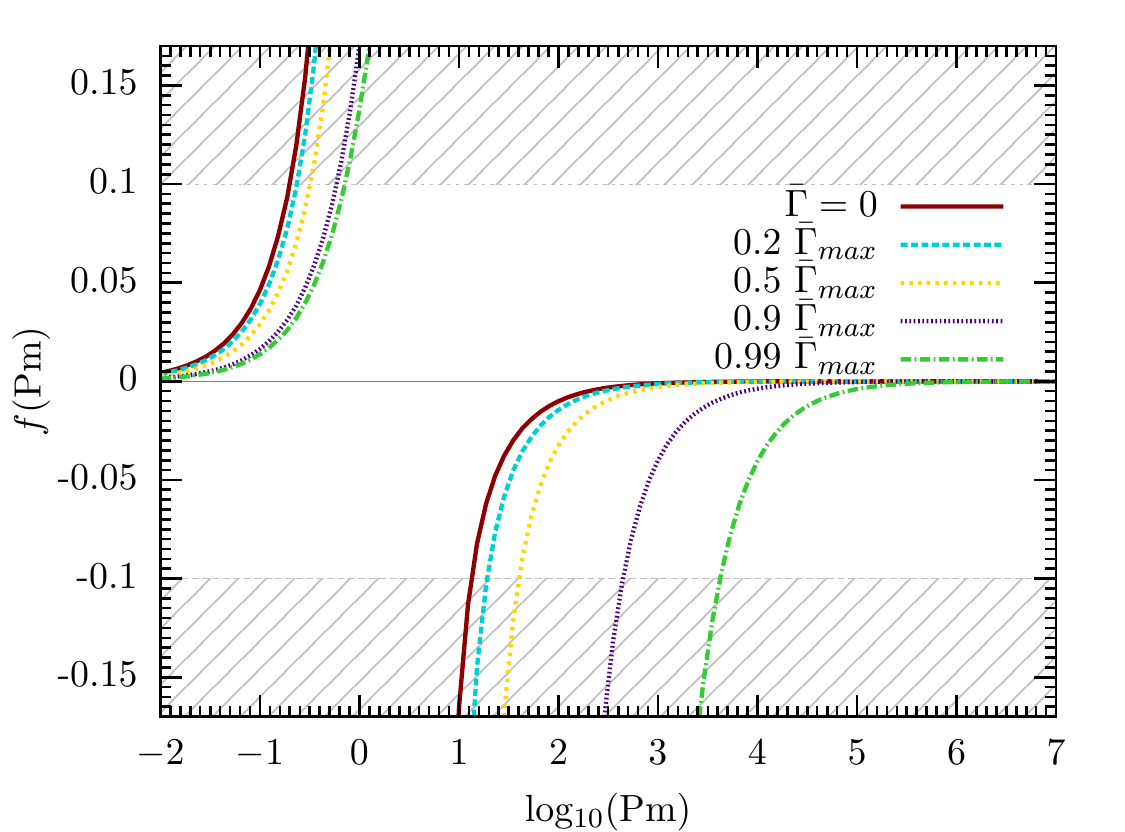}
  \caption{(Color online) The function $f(\text{Pm},\bar{\Gamma})$ for different values of the normalized growth rate for Kolmogorov turbulence.                            $\bar{\Gamma}_\text{max}$ is the normalized growth rate in the limit of infinite magnetic Prandtl numbers,                                 $\bar{\Gamma}_\text{max}=(37/36)\text{Re}^{1/2}$ (see Sec.~\ref{sec: Results} for the derivation). The WKB approximation                       is valid within the non-hatched area, i.e., for $\left|f(\text{Pm},\Gamma)\right|<0.1$.}
  \label{fig7}
\end{figure}

In order to check also lower Prandtl numbers we plot $f(\text{Pm},\Gamma)$ for different normalized growth rates $\bar{\Gamma}$ (\ref{normGamma}) and Kolmogorov turbulence in Fig.~\ref{fig7}. However, one can show that $f(\text{Pm},\Gamma)$ does not depend on the Reynolds number for Kolmogorov turbulence. So we choose values for $\bar{\Gamma}$ between $0$ and the maximal value $\bar{\Gamma}_\text{max}$ for the plot in Fig.~\ref{fig7}, where $\bar{\Gamma}_\text{max}$ is the value for an infinite Prandtl number and depends on the Reynolds number. One can see that the critical Prandtl number for the WKB approximation gets larger with increasing normalized growth rate.\\
To make a more quantitative estimate of the critical Prandtl number, we have hatched the area above $f(\text{Pm},\bar{\Gamma})=0.1$ and below $f(\text{Pm},\bar{\Gamma})=-0.1$. When $f$ is not in this area its absolute value is smaller than 10\% of 1. We take this as a threshold for our approximation.\\
We find that our method is applicable in the case of $\bar{\Gamma}=0$ for
\begin{equation}
  \text{Pm} \gtrsim 13.
\end{equation}
For higher normalized growth rates the critical Prandtl number increases.

\subsection{Validity of the WKB approximation for Burgers turbulence}

\begin{figure}[ht!]
    \centering
    \includegraphics[width=0.48\textwidth]{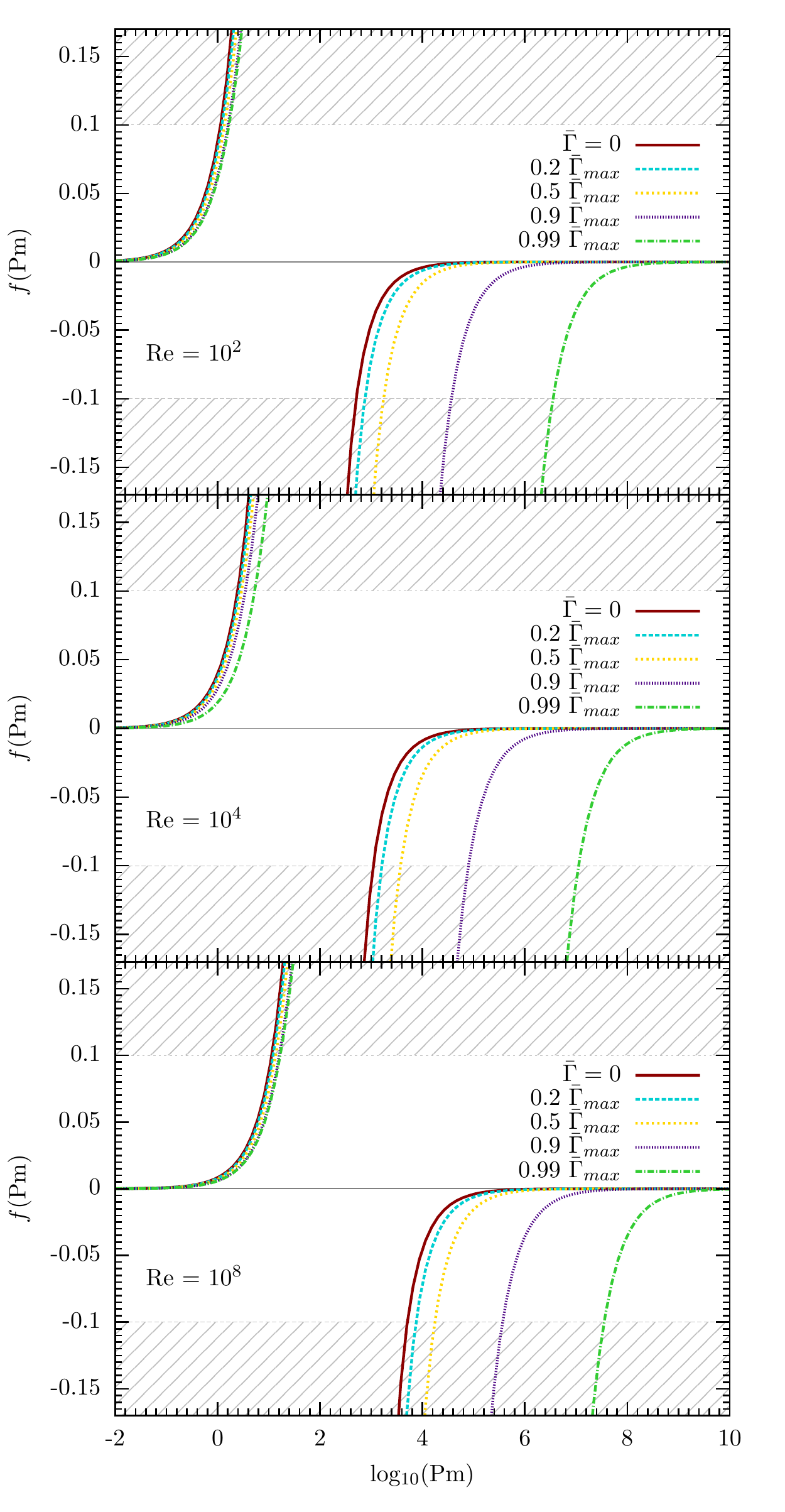}
    \caption{(Color online) The function $f(\text{Pm},\bar{\Gamma})$ for fixed Reynolds numbers and different values of the normalized growth rate for Burgers                turbulence.  Notice that in the limit $\text{Pm}\rightarrow\infty$ the normalized growth rates are $\bar{\Gamma}=0.85$ for                        $\text{Re}=10^{2}$, $\bar{\Gamma}=3.95$ for $\text{Re}=10^{4}$, and $\bar{\Gamma}=85.1$ for $\text{Re}=10^{8}$. The WKB approximation is valid                   within the non-hatched area, i.e.,  for $f(\text{Pm},\Gamma)<0.1$.}
    \label{fig8}
\end{figure}

We can analyze the validity of the WKB solutions for Burgers turbulence in the same way as for Kolmogorov turbulence using criterion (\ref{CheckApprox}).\\
However, we find that the function $f$ given in (\ref{f}) now depends not only on the normalized growth rate $\bar{\Gamma}$ and the Prandtl number Pm, but also on the Reynolds number Re. The result is shown in Fig.~ \ref{fig8}, where we plot $f$ against the Prandtl number for different Reynolds numbers and different normalized growth rates.\\
We again determine the critical Prandtl number for the WKB method for a vanishing normalized growth rate. For our different values of the Reynolds number we get the following critical Prandtl numbers at vanishing growth rate:
\begin{eqnarray}
  \text{Pm}(\text{Re}=10^2) & \gtrsim & 500, \\
  \text{Pm}(\text{Re}=10^4) & \gtrsim & 1100, \\
  \text{Pm}(\text{Re}=10^8) & \gtrsim & 5100.
\end{eqnarray}

\end{document}